\documentclass[aps,twocolumn,pra,groupedaddress,nofootinbib]{revtex4}

\usepackage{graphicx,amsmath,amsthm,amssymb,amsbsy,subfigure,hyperref,bbm,times,txfonts,float}
\usepackage{subfigure,hyperref,bbm,times}
\usepackage[T1]{fontenc}
\usepackage{braket}
\usepackage{epsfig}
\usepackage{color}
\usepackage{graphicx}% Include figure files
\usepackage{dcolumn}% Align table columns on decimal point
\usepackage{bm}% bold math

\providecommand{\openone}{\leavevmode\hbox{\small1\kern-3.8pt\normalsize1}}

\newtheorem*{rsl}{Main Result}

\usepackage{soul}

\hypersetup{
	colorlinks=true,
	linkcolor=blue,
}
\graphicspath{{figure/}}

\begin{document}
	
	%\preprint{APS/123-QED}
	
	\title{Hilbert-Schmidt speed as an efficient tool in quantum metrology}% Force line breaks with \\

	\author{Hossein Rangani Jahromi}
	\email{h.ranganijahromi@jahromu.ac.ir}
	\affiliation{Physics Department, Faculty of Sciences, Jahrom University, P.B. 74135111, Jahrom, Iran}
	
\author{Rosario Lo Franco}%
\email{rosario.lofranco@unipa.it}
\affiliation{Dipartimento di Ingegneria, Universit\`{a} di Palermo, Viale delle Scienze, Edificio 6, 90128 Palermo, Italy}%

	\date{\today}% It is always \today, today,

\begin{abstract}
We investigate how the Hilbert-Schmidt speed (HSS), a special type of
quantum statistical speed, can be exploited as  a powerful and easily computable tool for quantum phase estimation in a $n$-qubit system. We find that, when both the HSS and quantum Fisher information (QFI) are computed with respect to the phase parameter encoded into the initial state of the $n$-qubit register,    the zeros of the HSS dynamics are essentially the same
as those of the QFI dynamics. Moreover,  the positivity (negativity) of the time-derivative of the HSS exactly coincides with the positivity (negativity) of the time-derivative of the QFI. Our results also provide strong evidence for contractivity of the HSS under memoryless dynamics and its sensitivity to system-environment information backflows to detect the non-Markovianity in high-dimensional systems, as predicted in previous studies. 
%[Phys. Rev. A \textbf{102}, 022221 (2020)]. 
\end{abstract}

%Use showkeys class option if keyword
\maketitle

%\tableofcontents

\section{Introduction \label{introduction}}

When intending to estimate an unknown parameter in a quantum process, typically we  prepare the  probe system  in an initial state, let it interact with an environment to encode the information about the unknown parameter, and then measure the probe to extract the information and estimate the parameter. It should be noted that  in this process, the system also may be affected by different noises. 
Provided that the physical mechanism governing the system dynamics is known, we may
deduce an estimated value of the parameter by comparison between the input
and the output states of the probe \cite{giovannetti2006quantum}.

Phase estimation is at the heart of quantum
metrology  \cite{holevo1978estimation,giovannettiScience,paris2009quantum,giovannetti2011advances,liu2019quantum,
toth2014quantum,jafarzadeh2020effects,Jahromi2020,polinoReview,pirandolaReview,RevModPhys.90.035005} such that in many technological areas, the
estimation problem is concerned with determining  a phase shift of
the quantum state describing the probe. Estimation of  an unknown phase has many significant applications such as the
observation of gravitational waves \cite{PhysRevLett.116.061102} and detection of weak signals or defects resulting in the design of very sensitive sensors \cite{Taylor2013}. 
In most of these scenarios an interferometric scheme is  used to  implement the quantum phase estimation. The most important variations of the  interferometers include optical
interferometry in gravitational wave detectors, Ramsey spectroscopy in atomic physics, optical imaging or laser gyroscopes   to name but a few. All of these
applications usually aim at  optimal estimation of  a relative phase gathered by one \textit{arm } of the
interferometer \cite{dorner2009optimal}.
\par
According to the quantum Cram\'{e}r-Rao theorem, the precision of the quantum phase estimation is bounded by the inverse of the quantum Fisher information (QFI) \cite{helstrom1976quantum,paris2009quantum},
which thus denotes a central quantity in quantum metrology. In fact,
evaluation of the QFI provides the ultimate quantum limits to precision and consequently a
general benchmark to assess quantum metrological protocols.
\par
 The QFI is also  a 
measure of quantum statistical speed such that it quantifies the sensitivity of an initial state with respect to changes of the parameter
which should be estimated. 
The more sensitivity indicates that
the  parameter, which could be an unknown phase shift of
interest, can be estimated more efficiently,
or with more precision. On the other hand, each measure of statistical distance naturally leads to
a statistical speed for parametric evolutions of classical probability
distributions or quantum states. This statistical speed can be obtained
by the change in distance originated from a small change of this
parameter (i.e., the derivative of the distance). The quantum statistical speed is obtained by maximizing
over the classical statistical speed over all quantum measurements \cite{gessner2018statistical}. 

\par
Inspired by the fact that the QFI can be derived   as 
quantum statistical speed from the Hellinger distance \cite{jeffreys1946invariant}, we  investigate the application of the Hilbert-Schmidt speed (HSS), another   interesting quantifier of quantum statistical speed which has the advantage of avoiding diagonalization
of the evolved density matrix, in the quantum phase estimation. Because the computation of the QFI for high-dimensional quantum systems is very complicated, it would be useful to inquire the efficiency of the HSS, which is an easily computable quantity,  in the quantum estimation theory. 

In this paper, we show that the HSS can be exploited as a powerful and convenient figure of merit in quantum metrology for $n$-qubit systems. This result gains particular attention considering the fact that most of the quantum information protocols are designed by $n$-qubit registers.

The paper is organized as follows. In Sec.~\ref{Preliminaries} we briefly review the definition of the QFI and HSS. In Sec.~\ref{SECTIONMAIN} we present our main result about the applicability of the HSS in quantum phase estimation and check its validity by various examples. 
Finally, Sec.~\ref{cunclusion} summarizes the main results and prospects.

\section{ Preliminaries}\label{Preliminaries}

\subsection{ Quantum  Fisher information (QFI)}

We start by recalling the general formulation resulting in defining a kind of quantum statistical speed  by which the QFI  can be characterized. 
\par
First, we consider  the (classical) Hellinger
distance \cite{jeffreys1946invariant}

\begin{equation}\label{dis}
[d(p,q)]^{2}=\dfrac{1}{2}\sum\limits_{x}^{}|\sqrt{p_{x}}-\sqrt{q_{x}}|^{2},
\end{equation}
in which  $ p = \{p_{x}\}_{x} $ and $ q = \{q_{x}\}_{x} $ represent the probability distributions.
Here it has been assumed that the random variable $ x $  takes only  discrete
values.  

\par
Formally,
in order to achieve the statistical speed from a given statistical distance, one should
quantify the distance between infinitesimally close
distributions taken from a one-parameter family $ p_{x}(\varphi) $ with
parameter $ \varphi $. Following this prescription and performing a Taylor expansion at $ \varphi_{0} $ for small
values of $ \varphi $, we find that  the classical statistical speed associated with the (classical) Hellinger
distance  is given by 
\begin{equation}\label{classicalspeed}
s[p(\varphi_{0})]\equiv\dfrac{d}{d\varphi}d\big(p(\varphi_{0}+\varphi),p(\varphi_{0})\big)=\sqrt{\dfrac{f(p(\varphi_{0}))}{8}},
\end{equation}
where 
\begin{equation}\label{FI}
f(p(\varphi))=\sum_{x}^{}p_{x}(\varphi)\bigg(\dfrac{\partial~ \text{ln}p_{x}(\varphi)}{\partial \varphi}\bigg)^{2},
\end{equation}
denotes the Fisher information \cite{braunstein1994statistical,paris2009quantum}.
\par
Extending these classical notions to the quantum case with
considering  a given pair of quantum states $ \rho $ and $ \sigma $, one  
may write $ p_{x} = \text{Tr}\{E_{x}\rho\} $ and $ q_{x} = \text{Tr}\{E_{x}\sigma\} $ representing the measurement
probabilities associated with the positive-operator-valued measure (POVM) defined by the $ \{E_{x}\geq 0\} $ which satisfies $\sum\limits_{x}^{} E_{x} = \mathbb{I}  $.
The associated quantum distance can be obtained by maximizing the classical distance over all  possible choices of
POVMs \cite{PhysRevA.69.032106}, i.e., 
\begin{equation}\label{qdis}
D(\rho,\sigma)=\max_{\{E_{x}\}}d(p,q)=\sqrt{1-\mathcal{F}(\rho,\sigma)},
\end{equation}
called   the Bures distance \cite{braunstein1994statistical} in which $\mathcal{F}(\rho,\sigma)\equiv \text{Tr}\sqrt{\sqrt{\rho}\sigma \sqrt{\rho}}$ denotes the fidelity \cite{jozsa1994fidelity}.

Now we can define  the quantum statistical speed \cite{braunstein1994statistical} as follows

\begin{equation}\label{Qspeed}
S[\rho(\varphi_{0})]\equiv\dfrac{d}{d\varphi}D\big(\rho(\varphi_{0}+\varphi),\rho(\varphi_{0})\big)=\sqrt{\dfrac{F(\rho(\varphi_{0}))}{8}},
\end{equation}
where  the quantum Fisher information (QFI) is given by \cite{braunstein1994statistical,giovannetti2006quantum,giovannetti2011advances,liu2019quantum}
\begin{eqnarray}\label{QFII}
\nonumber F(\rho(\varphi))\equiv F_{\varphi}=\sum_{i,j}\frac{2}{\lambda_{i}+\lambda_{j}}|\langle \phi_{i}|\partial_{\varphi}\rho\left(\varphi \right)|\phi_{j}\rangle|^{2}~~~~~~~~~~~~~~~~~\\\
=\sum_{i} \frac{(\partial_{\varphi}\lambda_{i})^{2}}{\lambda_{i}}+2 \sum_{i\neq j} \frac{(\lambda_{i}-\lambda_{j})^{2}}{\lambda_{i}+\lambda_{j}} |\langle\phi_{i}|\partial_{\varphi} \phi_{j} \rangle|^{2},
\end{eqnarray} 
in which $ |\phi_{i}\rangle $ and $ \lambda_{i} $, respectively,  denote the eigenvectors and eigenvalues of  matrix $ \rho\left(\varphi \right) $. In fact, the QFI is obtained by maximizing the Fisher information over   all possible POVMs \cite{paris2009quantum}, i.e., 
\begin{equation}
F(\rho(\varphi))=\max_{\{E_{x}\}} f(p(\varphi)), 
\end{equation}
in which $p(\varphi) =\{p_{x}(\varphi)\}_{x}  $ and $p_{x}(\varphi) = \text{Tr}\{E_{x}\rho(\varphi)\}  $.
The fundamental relationship between the  QFI and its corresponding quantum
bound is expressed by the quantum  Cram\'{e}r-Rao bound stating that \cite{braunstein1994statistical}

\begin{equation}
\Delta \varphi_{QCR}=\sqrt{\dfrac{1}{F(\rho(\varphi))}}, 
\end{equation}
setting the precision limit  for quantum  estimation of unknown parameter $ \varphi $.

\subsection{Hilbert-Schmidt speed (HSS)}\label{sec:HSS}

Introducing the   distance measure \cite{gessner2018statistical}   
\begin{equation}\label{dis}
[\text{d}(p,q)]^{2}=\dfrac{1}{2}\sum\limits_{x}^{}|p_{x}-q_{x}|^{2},
\end{equation}
in which $ p = \{p_{x}\}_{x} $ as well as $ q = \{q_{x}\}_{x} $ are probability distributions,
 and
subsequently considering the classical statistical
speed
\begin{equation}\label{classicalspeed}
\text{s}\big[p(\varphi_{0})\big]=\dfrac{d}{d\varphi}\text{d}\big(p(\varphi_{0}+\varphi),p(\varphi_{0})\big),
\end{equation}
we can define a special kind of quantum statistical speed  which is called the HSS. 
Following the procedure discussed in the previous subsection for obtaining the corresponding quantum relations, one may obtain the Hilbert-Schmidt  distance \cite{ozawa2000entanglement} 
\begin{equation}\label{qdis}
\text{D}(\rho,\sigma)\equiv \max_{\{E_{x}\}}\text{d}(\rho,\sigma)=\sqrt{\frac{1}{2}\text{Tr}[(\rho-\sigma)^{2}]},
\end{equation}
not requiring the  diagonalization
of the argument operator. 
Moreover,  the corresponding quantum statistical speed also called  the HSS, is obtained as follows \cite{gessner2018statistical}
\begin{align}\label{HSSS}
HSS \bigg(\rho(\varphi)\bigg)\equiv HSS_{\varphi}
&\equiv \text{S}\big[\rho(\varphi)\big]=\max_{\{E_{x}\}} \text{s}\big[p(\varphi)\big]\nonumber\\
&=\sqrt{\frac{1}{2}\text{Tr}\bigg[\bigg(\dfrac{d\rho(\varphi)}{d\varphi}\bigg)^2\bigg]},
\end{align}
 which can be computed without  diagonalizing  $ \text{d}\rho(\varphi)/\text{d}\varphi $.

 \section{Quantum estimation through HSS}\label{SECTIONMAIN}
 
 Because both the QFI and  HSS are  quantum statistical speeds associated, respectively, with the Bures and Hilbert-Schmidt  distances, it is reasonable to investigate how they can be related to each other. 
By numerical simulation, we find that there is an important relationship between them. We state it in the following.

\begin{rsl}
	Suppose that we are given a  pure  initial state of a $n$-qubit quantum register, i.e.,
		\begin{equation}\label{initialasli}
	|\psi_{0}\rangle=N\sum_{j} \text{e}^{i\varphi_{j}}c_{j}|j\rangle
	\end{equation}
	in which $ N=\frac{1}{\sqrt{\sum_{j}|c_{j}|^{2}}} $ represents the normalization factor and 
	$ \left\lbrace |j\rangle \right\rbrace  $ denotes the computational basis. Then,  this state is affected by a general quantum channel $\mathcal{E}_{t}$ such that the output state  is given by $ \rho_{t} =\mathcal{E}_{t}(|\psi_{0}\rangle \langle \psi_{0} |)$. Under these conditions, we find that $ HSS_{\varphi_{j}}\equiv HSS \bigg(\rho_{t}(\varphi_{j})\bigg) $ and $F_{\varphi_{j}}\equiv F \bigg(\rho_{t}(\varphi_{j})\bigg)   $ computed with respect to phase parameter $ \varphi_{j} $ encoded into  
the input state (\ref{initialasli}),
	  exhibit qualitatively the same dynamics such that if   $ HSS_{\varphi_{j}} \neq 0   $, we have $\dfrac{\text{d}HSS_{\varphi_{j}}}{\text{d}t}\geq0 \Leftrightarrow \dfrac{\text{d}F_{\varphi_{j}}}{\text{d}t}\geq 0$ and $\dfrac{\text{d}HSS_{\varphi_{j}}}{\text{d}t}\leq0 \Leftrightarrow \dfrac{\text{d}F_{\varphi_{j}}}{\text{d}t}\leq0$. Moreover, $ HSS_{\varphi_{j}}=0 \Leftrightarrow F_{\varphi_{j}}=0 $. Therefore, investigating   the HSS dynamics, we can  detect the instants at which the optimal phase estimation is achieved. 
\end{rsl}
The  sanity check of this technique is performed by presenting  various examples in the following subsections. 
It should be noted that     using the  general hierarchy between the HSS and QFI discussed in  \cite{gessner2018statistical}, one can show that
$0 \leqslant HSS_{\varphi_{j}} \leqslant \sqrt{F_{\varphi_{j}}}$, hence  $ F_{\varphi_{j}} = 0 $ leads to $ HSS_{\varphi_{j}} $. However,   the reverse (i.e., detecting the QFI zeros through the HSS zeros,  which we discussed in this paper) cannot be necessarily extracted from the above general inequality.

\subsection{One-qubit example}\label{one-qubit example}
First we focus on  
 a one-qubit system  interacting  with a dissipative reservoir through the  Hamiltonian 
\begin{equation}\label{Ham}
H=\omega_{0}~\sigma_{+}\sigma_{-}+\sum\limits_{k}^{}\omega_{k}b^{\dagger}_{k}b_{k}+(\sigma_{+}B+\sigma_{-}B^{\dagger}),
\end{equation}
where $ \omega_{0} $ denotes the transition frequency of the qubit, $\sigma_{\pm}  $ represent the system raising and lowering operators, $ \omega_{k} $ is the frequency of the $k$-th field mode of the reservoir, $ b_{k} $ ($ b^{\dagger}_{k} $)  denotes the 
$k$-mode annihilation  (creation) operator, and $ B=\sum_{k}^{}g_{k}b_{k} $ in which $ g_{k} $ represents the coupling constant with the $k$-th mode.

At zero temperature and in the  strong-coupling regime using  Hamiltonian (\ref{Ham})  with a Lorentzian spectral density for the cavity modes and preparing the qubit in initial state 
\begin{equation}
|\psi_{0}\rangle=\text{cos}(\theta) |1\rangle+ \text{e}^{i\varphi}\text{sin}(\theta) |0\rangle,
\end{equation}
one  can find that the dynamics of the qubit  in basis $ \{|1\rangle,~|0\rangle\} $ is described  by the following evolved reduced density matrix \cite{breuer2002theory,bellomo2007non}
\begin{equation}\label{redudedq1}
\rho^{S}(t)=\left( \begin {array}{cc} \dfrac{P_{t} }{2} \left( \cos \left( \theta \right) +1
\right)
&\dfrac{\sqrt {P_{t} }}{2}{{\rm e}^{-i\varphi }}\sin \left( \theta
\right)
\\ \noalign{\medskip} \dfrac{\sqrt {P_{t} }}{2}{{\rm e}^{i\varphi }}\sin \left( \theta
\right)&1-\dfrac{P_{t} }{2} \left( \cos \left( \theta \right) +1
\right)
\end {array} \right),
\end{equation}
in which the coherence characteristic function $P(t)$ is
\begin{equation}\label{Pt}
P(t)=\text{e}^{-\lambda t}\left[\cos(\Gamma t/2)+(\lambda/\Gamma)\sin(\Gamma t/2)\right]^{2},
\end{equation}
with $ \Gamma=\sqrt{2\gamma_{0}\lambda-\lambda^{2}} $. The parameter $ \lambda $, connected to the reservoir correlation time $ \tau_{c} $ by  relation $ \tau_{c} \approx 1/\lambda$, represents the spectral width for the qubit-reservoir coupling. Moreover,  decay rate $\gamma_{0}  $ is related to the system (qubit) relaxation time scale $\tau_{r}   $, over which the state
of the system changes, by  relation  $ \tau_{r} =1/\gamma_{0}$. 

Inserting (\ref{redudedq1}) into Eqs. (\ref{QFII}) and (\ref{HSSS}), we find that the QFI and HSS associated with initial phase $ \varphi $, respectively, are given by
\begin{equation}\label{QFIq1}
F_{\varphi}(t)=P_{t}  \sin^{2} \left( \theta \right),
\end{equation}
and 
\begin{equation}\label{HSSq1}
HSS_{\varphi}(t)=\dfrac{\sqrt{P_{t}}}{2}\sin \left( \theta \right),
\end{equation}
leading to relations 
\begin{equation}\label{QFIHSSQ1}
 F_{\varphi}=4(HSS_{\varphi})^{2}\Longrightarrow\dfrac{\text{d}F_{\varphi}}{\text{d}t} =(8~HSS_{\varphi})\dfrac{\text{d}HSS_{\varphi}}{\text{d}t}.
\end{equation}
Accordingly, we see that when the HSS vanishes, the QFI also equals zero. Moreover,  at all instances when  $HSS_{\varphi} \neq 0 $, the signs of $  \dfrac{\text{d}F_{\varphi}}{\text{d}t} $ and $ \dfrac{\text{d}HSS_{\varphi}}{\text{d}t} $  are similar and hence they exhibit qualitatively the same dynamics. In particular,  the times at which the optimal estimation is achieved, i.e., $ \dfrac{\text{d}F_{\varphi}}{\text{d}t} =0 $, can be easily detected by investigating the HSS dynamics.

\subsection{Two-qubit examples}
Here the validity of our result for two-qubit systems in three different scenarios, i.e., coupling to  independent environments, interaction with common environment, and teleportation of the entanglement between the qubits, is discussed.

\subsubsection{Coupling to  independent environments}
We now study a composite quantum system which consists
of two separated qubits  independently interacting with their own dissipative reservoir.
Knowing the evolved density matrix of the single qubit discussed in previous subsection, one can easily obtain the density matrix evolution of the two independent qubits \cite{bellomo2007non}. We investigate the scenario in which  the two qubits are prepared in  initial state 
\begin{equation}\label{initial2}
|\psi_{0}\rangle=\frac{1}{\sqrt{3}}(\text{e}^{i\varphi}|10\rangle+|01\rangle+|00\rangle),
\end{equation}
resulting in the  evolved reduced density matrix 
 \begin{equation}
 \rho^{S}(t)=\left(
\begin{array}{cccc}
0 & 0 & 0 & 0 \\
0 & \dfrac{P_{t}}{3} & \dfrac{P_{t}}{3} e^{i \varphi }  & \dfrac{\sqrt{P_{t}}}{3} e^{i \varphi }  \\
0 & \dfrac{P_{t}}{3} e^{-i \varphi }  & \dfrac{P_{t}}{3} & \dfrac{\sqrt{P_{t}}}{3} \\
0 & \dfrac{\sqrt{P_{t}} }{3} e^{-i \varphi } & \dfrac{\sqrt{P_{t}}}{3} & 1-\dfrac{2 P_{t}}{3} \\
\end{array}
\right),
  \end{equation}
  where $P_{t}\in[0,1]$ is the coherence characteristic function of Eq.~(\ref{Pt}). 
Computing the QFI and HSS with respect to phase parameter $ \varphi $, one promptly gets that they are given, respectively, by
\begin{equation}\label{QFIq2}
F_{\varphi}(t)=\dfrac{8}{9}P_{t},\quad
HSS_{\varphi}(t)=\frac{1}{3} \sqrt{P_{t} (P_{t}+1)}.
\end{equation}
resulting in  
\begin{eqnarray}\label{QFIHSSQ2}
  F_{\varphi}&=&\dfrac{4}{9}(\,\sqrt {1+36\,{{\it HSS^{2}_{\varphi}}}}-1) 
  \Rightarrow \nonumber\\
  \dfrac{\text{d}F_{\varphi}}{\text{d}t} &=& {\frac {{\it 16~ HSS_{\varphi}}}{\sqrt {1+36\,{{\it HSS^{2}_{\varphi}}}}}}  \dfrac{\text{d}HSS_{\varphi}}{\text{d}t}. 
 \end{eqnarray}
Again assuming that $ HSS_{\varphi_{j}} \neq 0   $, we have $\dfrac{\text{d}HSS_{\varphi_{j}}}{\text{d}t}\geq0 \Leftrightarrow \dfrac{\text{d}F_{\varphi_{j}}}{\text{d}t}\geq 0$ and $\dfrac{\text{d}HSS_{\varphi_{j}}}{\text{d}t}\leq0 \Leftrightarrow \dfrac{\text{d}F_{\varphi_{j}}}{\text{d}t}\leq0$. In addition, at instants when $ HSS_{\varphi_{j}}=0$, the QFI also vanishes and hence no information can be extracted from the system.

\subsubsection{Interaction with common environment}

\begin{figure} [t!]
	\centering
	\includegraphics[width=0.45\textwidth]{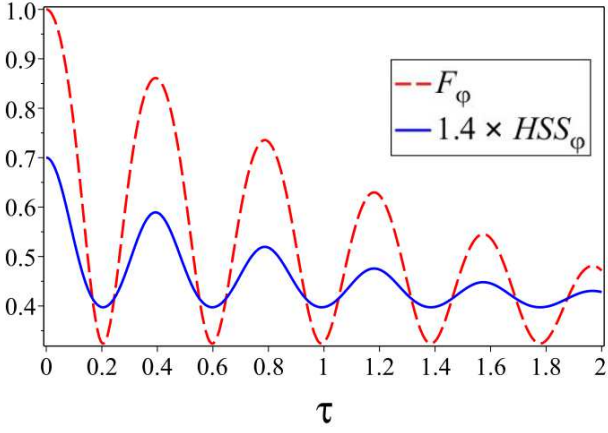}
	\caption{Dynamics of quantum Fisher information  $F_{\varphi}(t) $ (red dashed line) and  Hilbert-Schmidt speed $HSS_{\varphi}(t) $ (amplified by $1.4$ times for comparison, blue solid line),   as a function of the dimensionless time $\tau$ for the two-qubits system coupled to a
		common reservoir, with $ r_{1}=0.3$ and $ R=8 $. }\label{plotcommon}
\end{figure}

We study two qubits interacting with a
common zero-temperature bosonic reservoir. The  total Hamiltonian of the two-qubit system plus the reservoir is written as $ H=H_{0} +H_{int}$, with \cite{PhysRevLett.100.090503}
\begin{eqnarray}
	H_{0}&=&\omega_{1}~\sigma^{(1)}_{+}\sigma^{(1)}_{-}+\omega_{2}~\sigma^{(2)}_{+}\sigma^{(2)}_{-}+\sum\limits_{k}^{}\omega_{k}b^{\dagger}_{k}b_{k},\nonumber\\
H_\mathrm{int}&=&(\alpha_{1}\sigma^{(1)}_{+}+\alpha_{2}\sigma^{(2)}_{+})B+(\alpha_{1}\sigma^{(1)}_{-}+\alpha_{2}\sigma^{(2)}_{-})B^{\dagger},
\end{eqnarray}
where  $\sigma^{(j)}_{\pm}  $ and  $ \omega_{j} $  denote, respectively, the inversion operator and transition frequency of the $ j $th qubit, $ j=1,2 $, $ b^{\dagger}_{k} $ ($b_{k}  $)  represents the 
	$k$-mode creation  (annihilation) operator of quanta of the
	environment, and $ B=\sum_{k}^{}g_{k}b_{k} $ in which $ g_{k} $ is the coupling constant with the $k$-th mode. Moreover, the interaction of the $ j $th qubit with the reservoir is measured by the dimensionless constant $ \alpha_{j} $ depending on the value of the
	cavity field at the qubit position and can be effectively
	controlled by means of dc Stark shifts tuning the
	atomic transition in and out of the resonance. We investigate the case in which the two
	atomic qubits interact resonantly with the reservoir described by a Lorentzian
	spectral density
	 and they have the same transition frequency, i.e., $ \omega_{1} =\omega_{2} = \omega_{0}  $.
	 \par
	 It is useful to introduce a collective
	 coupling constant $ \alpha_{T}=\sqrt{\alpha^{2}_{1}+\alpha^{2}_{2}} $, the relative
	 strengths $ r_{j}=\alpha_{j}/\alpha_{T} $ such that  $r_{1}^{2}+r_{2}^{2}=1  $, and  
	 mutually orthogonal quantum states
	  \begin{equation}
	  | \psi_{ +}\rangle  =r_{1}| 10\rangle+r_{2}| 01\rangle,\quad
| \psi_{ -}\rangle  =r_{2}| 10\rangle-r_{1}| 01\rangle.
	   \end{equation}
	   With these definitions, one finds that for an initial state of the form 
	    \begin{equation}
	  | \psi_{ 0}\rangle  =\big[\dfrac{1}{\sqrt{2}}(| 10\rangle+\text{e}^{i\varphi}| 01\rangle)\big]\bigotimes_{k}  | 0_{k}\rangle,
	  \end{equation}
and  in the basis $ \{|1\rangle,|0\rangle\} $, the reduced density operator for the
the two-qubit system is written as \cite{PhysRevLett.100.090503}
\begin{equation}\label{reducedcommon}
\rho(t)=\left(
\begin{array}{cccc}
0& 0&0&0  \\
0 &|c_{1}(t)|^{2}&c_{1}(t)c^{*}_{2}(t)&0  \\
0 &c^{*}_{1}(t)c_{2}(t) &|c_{2}(t)|^{2}&0  \\
0 & 0&0&1-|c_{1}(t)|^{2}-|c_{2}(t)|^{2}  \\
\end{array} \right),
\end{equation}
where, considering $ \beta_{\pm}= \langle \psi_{ \pm} | \psi_{ 0}\rangle  $, one has
 \begin{equation}
  c_{1}(t)=r_{2}\beta_{-}+r_{1}\mathcal{F}_{t}\beta_{+},\quad
c_{2}(t)=-r_{1}\beta_{-}+r_{2}\mathcal{F}_{t}\beta_{+}.
\end{equation}
Moreover,  defining dimensionless quantities $ \tau=\lambda t $ and $ R=\mathcal{R}/\lambda $ in which $ 1/\lambda $ is the  reservoir correlation
time and $ \mathcal{R} $ denotes the vacuum Rabi frequency, we have 
\begin{equation}
\mathcal{F}_{\tau}=\text{e}^{-\tau/2}\left[\cosh(\frac{\tau}{2}\sqrt{1-4R^{2}})+\dfrac{1}{\sqrt{1-4R^{2}}}\sinh(\frac{\tau}{2}\sqrt{1-4R^{2}})\right].
\end{equation}

The QFI can be computed analytically, however it is too complicated to present here. On the other hand, we find that the HSS is given by
\begin{widetext}
\begin{equation}
HSS_{\varphi}(t)=\dfrac{1}{2} \left( {r_{{1}}}^{2}+{r_{{2}}}^{2} \right) \sqrt {{\mathcal{F}_{t}}^{2}
	{r_{{1}}}^{4}-2\, \left( {\mathcal{F}_{t}}^{2}-1 \right) ^{2}{r_{{1}}}^{2}{r_
		{{2}}}^{2}\cos \left( 2\,\varphi  \right) +2\, \left( {\mathcal{F}_{t}}^{4}-{
		\mathcal{F}_{t}}^{2}+1 \right) {r_{{1}}}^{2}{r_{{2}}}^{2}+{\mathcal{F}_{t}}^{2}{r_{{
				2}}}^{4}}.
\end{equation}
\end{widetext}

Figure \ref{plotcommon}  illustrates that the QFI and HSS dynamics simultaneously exhibit an oscillatory behavior such that their maximum and minimum points exactly coincide. This figure qualitatively verifies our result that the HSS can detect exactly the times at which the best phase estimation occurs. In fact, the HSS, similar to QFI,  can be used as a distinguishability metric on the space of quantum states
which quantifies the maximum amount of information on  an unknown phase parameter
attainable by a given probe state.

\subsubsection{Two-qubit teleportation }
   One of the most  important models used in
the low-temperature regime is typically the spin environment \cite{jahromi2019multiparameter,Fernando}. In particular, in order to achieve  the proper operation in experiments performed to study the  macroscopic quantum coherence and decoherence, one require temperatures close to absolute zero.
Here we  consider a two-qubit system interacting with an external  environment composed of $ N $ spins. The general Hamiltonian  is therefore written as $ H=H_{S}+H_{E}+H_{I} $ where
the system, environment and  interaction Hamiltonians, respectively, are given by
\begin{equation}\label{HS2}
H_{S}=\dfrac{\hbar \Omega_{1}}{2}\sigma_{z}^{1}+\dfrac{\hbar \Omega_{2}}{2}\sigma_{z}^{2}+\gamma \sigma_{z}^{1}\sigma_{z}^{2},
\end{equation}
\begin{equation}\label{HSI2}
H_{E}=\sum\limits_{i=1}^{N} h_{i} \sigma_{xi},
\end{equation}
\begin{equation}\label{HI2}
H_{I}=\sigma_{z}^{1}\otimes \sum\limits_{n=1}^{N}\varepsilon_{i}\sigma_{zi}+\sigma_{z}^{2}\otimes \sum\limits_{n=1}^{N}\lambda_{i}\sigma_{zi},
\end{equation}
in which  $ \Omega_{i} $ and $ \gamma $  denote, respectively, the characteristic frequency of $ i $th qubit, and  the coupling strength between
the two spin qubits.
Moreover,
 $ \xi_{i} $ ($ \lambda_{i} $)  represents  the coupling
between qubit 1 ( qubit 2)  and the spins of the environment, and $ h_{i} $  denotes the tunneling matrix element
for the $ i $th-environmental spin.

Preparing the two-qubit system in  initial state 
\begin{equation}\label{INITIALtel}
\rho(0)=\frac{1-r}{4}\mathcal{I}+r|\vartheta\rangle \langle \vartheta|
\end{equation}
where $ r \in (0,1] $ denotes the mixing of the state, $\mathcal{I}  $ is $ 4\times 4 $ unity operator and 
\begin{equation}\label{BELL}
| \vartheta\rangle=\sqrt{1-p}|00\rangle+\sqrt{p}|11\rangle; ~~0\leq p \leq 1,
\end{equation}
one finds that  the evolved reduced density matrix is given by \cite{Fernando}
\begin{widetext}
\begin{equation}\label{reducedtel}
\rho(t)=\left(
\begin{array}{cccc}
\dfrac{1-r}{4}+r(1-p)& 0&0&r\sqrt{p(1-p)}e^{-i(\Omega_{1}+\Omega_{2})t} Q(t)  \\
0 &\dfrac{1-r}{4}&0&0  \\
0 &0 &\dfrac{1-r}{4}&0  \\
r\sqrt{p(1-p)}e^{i(\Omega_{1}+\Omega_{2})t} Q(t)  & 0&0&\dfrac{1-r}{4}+rp   \\
\end{array} \right),
\end{equation}
\end{widetext}
where the decoherence factor $ Q(t) $ is 

\begin{equation}\label{Q}
Q(t)=\prod\limits_{i=1}^{N}\bigg\lgroup  1-\big[\dfrac{2(\varepsilon_{i}+\lambda_{i})^{2}}{h_{i}^{2}+(\varepsilon_{i}+\lambda_{i})^{2}}\big] \text{sin}^{2}(t\sqrt{h_{i}^{2}+(\varepsilon_{i}+\lambda_{i})^{2}})\bigg\rgroup.
\end{equation}

Assuming that the two qubits are shared between Alice and Bob, 
we use  two copies of this system as a resource   for teleportation 
 of an unknown entangled state $ \rho_{in} $.
It is useful to introduce the Bell states  $ \mathcal{B}_{i} $'s associated with the Pauli matrices $ \sigma_{i} $'s by
 \begin{equation}\label{a2}
 \mathcal{B}_{i}=\left(\sigma_{0}\otimes\sigma_{i}\right)\mathcal{B}_{0}\left(\sigma_{0}\otimes\sigma_{i}\right); \;i=1,2,3, 
 \end{equation}
 in which $ \sigma_{0}=I $, $ \sigma_{1}=\sigma_{x} $, $ \sigma_{2}=\sigma_{y} $, $ \sigma_{3}=\sigma_{z} $, and $I$ represents the $2\times 2$ identity matrix.
In addition,   we choose $ \mathcal{B}_{0}=\frac{1}{2}\left(\Ket{00} +\Ket{11}\right) \left(\Bra{00} +\Bra{11}\right)$
 where  $\left\lbrace \Ket{0}, \Ket{1}\right\rbrace   $ is the usual computational basis for   the one-qubit system. Now,
 following  \cite{PhysRevLett.84.4236}, one can  
 generalize the standard teleportation protocol $ T_{0} $  and find that
 the output state of the two-qubit teleportation is given by \cite{PhysRevLett.87.267901}
\begin{equation}\label{a3}
\rho_{\text{out}}=\sum_{ij}p_{ij}\left( \sigma_{i}\otimes\sigma_{j}\right) \rho_{\text{in}}\left( \sigma_{i}\otimes\sigma_{j}\right),\  i,j=0,x,y,z,
\end{equation}
where
 $
  p_{ij}=\text{Tr}\left(\mathcal{B}_{i}\rho_{\text{res}} \right)\text{Tr}\left(\mathcal{B}_{j}\rho_{\text{res}} \right)
  $
  in which $\rho_\text{res}$, the resource state for the teleportation, equals the reduced density matrix (\ref{reducedtel}) in our model. Accordingly, for the input state $ \rho_{in}=\Ket{\psi_{ in}}\Bra{\psi_{ in}} $ with 
  \begin{equation}\label{inputtel}
  \Ket{\psi_{ in}}=\text{cos}(\theta/2)\Ket{10}+\text{sin}(\theta/2)\text{e}^{i\varphi}\Ket{01},
  \end{equation}
  where $ \ 0\leq\theta\leq\pi,\ 0\leq\varphi\leq2\pi $,
   we find that  the output state of the teleportation can  be
  written as
  \begin{widetext}
  	\begin{equation}\label{rhoouttel}
 \rho_{out}(t)=\left[ \begin {array}{cccc} 4\,{R}^{2}+2\,Rr&0&0&0
 \\ \noalign{\medskip}0& \left( 4\,Rr+{r}^{2} \right)   \sin^{2}
 \left( \dfrac{\theta}{2} \right)   +4\,{R}^{2}&2\,{e}^{i\varphi }
 \sin \left( \theta \right)  {A^{2}(t)}
 \cos^{2} \big( \Omega_{{1}}+\Omega_{{2}} \big)   &0
 \\ \noalign{\medskip}0&2\,{e}^{-i\varphi }
 \sin \left( \theta \right)  {A^{2}(t)}
 \cos^{2} \big( \Omega_{{1}}+\Omega_{{2}} \big) & \left( 4\,Rr+{r}^{2}
 \right)  \cos^{2} \left(\dfrac{ \theta}{2} \right)  +4\,{R}^{2}
 &0\\ \noalign{\medskip}0&0&0&4\,{R}^{2}+2\,Rr\end {array} \right],
  \end{equation}
\end{widetext}
  in which $ R=\dfrac{1-r}{4} $ and $ A(t)=\sqrt { \left( 1-p \right) p}r Q(t)   $.
  Using this expression for the output state,   we find that the  QFI and HSS associated to the phase parameter encoded into the input state used in teleportation channel are given, respectively, by
   \begin{equation}\label{QFITEL}
   F_{\varphi}(t)=32\,{\frac {{A}^{4}(t)  \cos^{4}  \big( \Omega_{{1}}+\Omega_{{2}}
   		\big)   \sin ^{2} \theta  
   	}{1+{r}^{2}}},
   \end{equation}
    \begin{equation}\label{HSSTEL}
  HSS_{\varphi}(t)=2\,{A}^{2}(t)  \cos^{2} \big( \Omega_{{1}}+\Omega_{{2}} \big) 
   \sin \theta.
   \end{equation}
   Therefore, we obtain 
    \begin{equation}\label{HSSQFITEL}
    F_{\varphi}={\frac {{{\it 8~HSS^{2}_{\varphi}}}}{1+{r}^{2}}} \Longrightarrow\dfrac{\text{d}F_{\varphi}}{\text{d}t}={\frac {{{\it 16~HSS_{\varphi}}}}{1+{r}^{2}}}\dfrac{\text{d}HSS_{\varphi}}{\text{d}t},
   \end{equation}
 leading to our main result, i.e., the possibility of extracting the QFI dynamics  through the HSS dynamics.

  \subsection{ $n$-qubit example ($ n\geq 3 $)}
  
   First we consider the dynamics of a topological qubit realized by two Majorana modes which are generated at the endpoints of some nanowire with strong spin-orbit
  interaction, placed on top of an s-wave superconductor and driven by an external magnetic field \textbf{B}  along the wire axis direction \cite{ho2014decoherence,jahromi2019quantum}. We also assume that each  Majorana mode  is coupled to the metallic nanowire via a tunnel junction  in the way that the  tunneling strength is controllable by an external gate voltage. 

  \par
  The total Hamiltonian is written as
  \begin{equation}
  H=H_{S}+H_{E}+V 
  \end{equation}
  in which $ H_{s} $ denotes the Hamiltonian of the topological qubit  and $ V $ represents the system-environment interaction Hamiltonian. In addition, the  environment Hamiltonian is denoted by $ H_{E} $ whose elementary
  constituents can be considered as electrons or holes. The decoherence which affects  the topological qubit  is modelled as a fermionic
  Ohmic-like environment described by spectral density $ \rho_{spec}\propto \omega^{Q} $ with $ Q\geq 0 $.   The  Ohmic, supere Ohmic and sub-Ohmic environments are characterized by $ Q = 1 $,  $ Q > 1 $ and 
  $ Q < 1 $, respectively.

 Since these Majorana modes used as the topological qubit are zero-energy modes, we have $ H_{S}=0 $.
  Moreover,  interaction Hamiltonian $ V $   constructed by the electron creation  (annihilation) operators with Majorana modes $ \gamma_{1} $  
  and $ \gamma_{2} $ satisfies the properties:
  \begin{equation}\label{MqubitProp}
  \gamma^{\dagger}_{a}=\gamma_{a},~~~\{\gamma_{a},\gamma_{b}\}=2\delta_{ab},
  \end{equation}
  where $ a,b=1,2 $. Before  turning on  interaction $ V $, the two
  Majorana modes construct a topological (non-local) qubit with states $ |0\rangle $ and $ |1\rangle $ related to each other by 
  \begin{equation}\label{Majorqubit}
  \frac{1}{2}(\gamma_{1}-\text{i}\gamma_{2})|0\rangle=|1\rangle,~~~~~\frac{1}{2}(\gamma_{1}+\text{i}\gamma_{2})|1\rangle=|0\rangle,
  \end{equation}
  where   the following representation has been chosen  for $ \gamma_{1,2} $:
  \begin{equation}\label{MajorPauli}
  \gamma_{1}=\sigma_{1},~~~\gamma_{2}=\sigma_{2},~~~\text{i}\gamma_{1}\gamma_{2}=\sigma_{3},
  \end{equation}
  in which $ \sigma_{j} $'s represent the Pauli matrices.
  
  Assuming that
  $ \varrho^{T} $, denoting the state of the total system, is uncorrelated initially: $\varrho^{T}(0)=\varrho(0)\otimes \varrho_{E}$,
  in which $\rho_{S}(0)$ and $\rho_{E}  $ are the initial density matrices of the topological qubit and its environment, respectively. Supposing that the initial state of the Majorana qubit is written as
  \begin{equation}\label{InitialMajorqubit}
  \varrho(0)=\left(\begin{array}{cc}
  \varrho_{11}(0)&\varrho_{12}(0)  \\
  \varrho_{21}(0)&\varrho_{22}(0)  \\
  \end{array}\right),
  \end{equation}
 one finds that the reduced
  density matrix of the system at time $ t $ can be obtained by dynamical map $ \Phi_{t} $ such that  (for details, see \cite{ho2014decoherence}):
   \begin{align}\label{Reduced1qMajorqubit}
  \varrho(t)&=\Phi_{t}\big(\varrho(0)\big)=\nonumber\\
  &\frac{1}{2}\left(\begin{array}{cc}
  1+(2\varrho_{11}(0)-1)\alpha^{2}(t)&2\varrho_{12}(0)\alpha(t)  \\
  2\varrho_{21}(0)\alpha(t)&1+(2\varrho_{22}(0)-1)\alpha^{2}(t) \\
  \end{array}\right),
  \end{align}
  in which 
  \begin{equation}\label{alpha}
  \alpha(t)=\text{e}^{-2B^{2}|\beta|I_{Q}(t)},~~~~~\beta\equiv \dfrac{-4\pi}{\Gamma(Q+1)}(\dfrac{1}{\varGamma_{0}})^{Q+1}
  \end{equation}
  where $ \varGamma_{0} $ denotes   the high-frequency cutoff for the linear spectrum of the edge state and $ \Gamma (z) $ represents the Gamma function. In addition,
  \begin{equation}\label{IQ}
  I_{Q}(t)= \left\{
  \begin{array}{rl}
  2\varGamma_{0}^{Q-1} \Gamma(\frac{Q-1}{2})\bigg[1-\,_1F_1\big(\frac{Q-1}{2};\frac{1}{2};-\frac{t^{2}\varGamma^{2}_{0}}{4}\big)\bigg], &  Q\neq 1,\\
  \frac{1}{2}t^{2}\varGamma^{2}_{0} \,_2F_2\bigg(\{1,1\};\{3/2,2\};-\frac{t^{2}\varGamma^{2}_{0}}{4}\bigg),   & Q=1,
  \end{array} \right.
  \end{equation}
where ${}_pF_q $ is the \textit{generalized hypergeometric function} and $ \varGamma(z) $ denotes  the Gamma function.
 
  From the eigenvalues and eigenvectors of  the \textit{Choi matrix} \cite{leung2003choi} of the map $ \Phi_{t} $,  the corresponding Kraus operators $\{K_{i}(t)\}  $ can be obtained as
    \begin{align}\label{Reduced1qMajorqubit}
   &K_{1}(t)=\left(
  \begin{array}{cc}
  \frac{\alpha -1}{2} & 0 \\
  0 & \frac{1-\alpha }{2} \\
  \end{array}
  \right),~K_{2}(t)=\left(
  \begin{array}{cc}
  \frac{\alpha +1}{2} & 0 \\
  0 & \frac{\alpha +1}{2} \\
  \end{array}
  \right),\nonumber\\
  &K_{3}(t)=\left(
  \begin{array}{cc}
  0 & \frac{\sqrt{1-\alpha ^2}}{\sqrt{2}} \\
  0 & 0 \\
  \end{array}
  \right),~K_{4}(t)=\left(
  \begin{array}{cc}
  0 & 0 \\
  \frac{\sqrt{1-\alpha ^2}}{\sqrt{2}} & 0 \\
  \end{array}
  \right).
  \end{align}
 
  Now, we consider a system formed by  $ n $ noninteracting topological qubits such that each qubit  locally interacts with the environment described above.  Note that the effects of the environment on each of the qubits can be canceled  by setting  the corresponding external magnetic field $ \textbf{B} $ 
  to zero. With this in mind  we focus on the scenarios in which $ m $ of the qubits are affected by the noise, while others  are noiseless. Since  the 
  environments are independent in our model, the Kraus operators are just tensor products of
  Kraus operators of each of the qubits, noting  that the Kraus operators of the noiseless qubits are set to identity operator.

  Using a three-qubit system $ (n=3) $  with $ m=3 $, initially prepared in the W-like state
  \begin{equation}\label{W}
  |\psi_{0}\rangle=\dfrac{1}{\sqrt{3}}\big(\text{e}^{i\varphi_{1}}|100\rangle+|010\rangle+\text{e}^{i\varphi_{2}}|001\rangle\big),
  \end{equation}
  we find that the QFI and HSS associated with phase parameter $ \varphi_{1} $ is obtained as
\begin{equation}\label{QFIHSS3}
F_{\varphi_{1}}(t)=\frac{2}{9} \left(5-\frac{2}{\alpha ^2(t)+1}\right),~HSS_{\varphi_{1}}(t)=\frac{\alpha ^2(t)+1}{3 \sqrt{2}}.
\end{equation}
  It is easily found that
  \begin{equation}\label{FHSSW}
  F_{\varphi_{1}}(t)={\frac {10}{9}}-{\frac {2\,\sqrt {2}}{27\,{\it HSS_{\varphi_{1}}}}}(t)\Rightarrow \dfrac{\text{d}F_{\varphi_{1}}}{\text{d}t}={\frac {2\,\sqrt {2}}{27\,{{\it (HSS_{\varphi_{1}}}})^{2}}} \dfrac{\text{d}HSS_{\varphi_{1}}}{\text{d}t}.
  \end{equation}
   As it is clear from  (\ref{QFIHSS3}), the HSS is always nonzero and therefore, according to  (\ref{FHSSW}),  we conclude that the QFI dynamics can be completely determined by analyzing the HSS dynamics, i.e., $\dfrac{\text{d}HSS_{\varphi_{1}}}{\text{d}t}> 0 \Leftrightarrow \dfrac{\text{d}F_{\varphi_{1}}}{\text{d}t}> 0$ and $\dfrac{\text{d}HSS_{\varphi_{1}}}{\text{d}t}<0 \Leftrightarrow \dfrac{\text{d}F_{\varphi_{1}}}{\text{d}t}<0$. 
   Now we compute the measures associated with phase parameter   $ \varphi_{2} $, leading to the  expressions
   \begin{equation}\label{QFI2HSS3}
   F_{\varphi_{2}}(t)=\frac{16 \alpha ^2(t)}{9 \alpha ^2(t)+9},\quad 
   HSS_{\varphi_{2}}(t)=\frac{\sqrt{2} \alpha(t) }{3}.
   \end{equation}
    Because 
 \begin{eqnarray}\label{F2HSSW}
    F_{\varphi_{2}}(t)&=&\dfrac{16~ \bigg(HSS_{\varphi_{2}}(t)\bigg)^{2}}{9~ \bigg(HSS_{\varphi_{2}}(t)\bigg)^{2}+2}\nonumber \\ 
    &\Rightarrow& \dfrac{\text{d}F_{\varphi_{2}}}{\text{d}t}= \,{\frac {{64~ HSS_{\varphi_{2}}}}{ \left[ 9\,(HSS_{\varphi_{2}}) ^{2}+2 \right] ^{2}}} \dfrac{\text{d}HSS_{\varphi_{2}}}{\text{d}t},
    \end{eqnarray}
   our main result can be again easily confirmed.

  As the final example, we take the $n$-qubit register prepared in a Greenberger-
  Horne-Zeilinger (GHZ)-like state \cite{greenberger1990bell} written as
  \begin{equation}\label{GHZNN}
  |\mathrm{GHZ}\rangle_{n}=\dfrac{1}{\sqrt{2}}\big(\text{e}^{i\varphi}|0\rangle^{\otimes n}+|1\rangle^{\otimes n}\big).
  \end{equation}
 Calculating the evolved state of the system, we find that the corresponding QFI and HSS are given, respectively, by 
 \begin{eqnarray}\label{QFIn}
 F^{GHZ}_{n,m}(\varphi)&=&F_{\varphi}=\bigg(\dfrac{2\alpha^{2}}{1+\alpha^{2}}\bigg)^{m},\nonumber\\
 HSS^{GHZ}_{n,m}(\varphi)&=&HSS_{\varphi}=\dfrac{\alpha^{m}}{2}.
 \end{eqnarray}
 Hence, we can write
  \begin{widetext}
 \begin{equation}\label{HSSQFIn}
 F_{\varphi}= 2^{m+2}(HSS_{\varphi})^2 \left(4^{1/m} (HSS_{\varphi})^{2/m}+1\right)^{-m}\Longrightarrow \dfrac{\text{d}F_{\varphi}}{\text{d}t}= 2^{m+3}(HSS_{\varphi}) \left(4^{1/m} (HSS_{\varphi})^{2/m}+1\right)^{-m-1}\dfrac{\text{d}HSS_{\varphi}}{\text{d}t},
 \end{equation}
  \end{widetext}
 explicitly leading to our main result, as described in other examples.

\section{Conclusions}\label{cunclusion}
	Quantum information processing based on $n$-qubit registers provides a playground for fundamental
research and also results in technological advances. Examples include stronger
violations of local realistic world views which can be used to tolerate larger
amounts of noise in quantum communication protocols.

In this paper, we have constructed a strong relationship between the Hilbert-Schmidt speed (HSS), which is a special
case of quantum statistical speed and the quantum Fisher information (QFI),
a key concept in parameter estimation theory, for $n$-qubit systems.
  The idea underlying this relationship stems from the fact that the QFI, quantifying  the sensitivity of an initial state with respect to changes of the parameter
of a dynamical evolution, is a quantum statistical speed extracted from the Hellinger distance. In contrast to the computational complication of the QFI, especially for multipartite systems, our findings show that the HSS can be instead employed as a strong and efficient tool in quantum metrology, because of its straightforward determination. 
%causes the QFI determination to be a complicated task for high-dimensional systems. 

The QFI  monotonically 
 decreases under Markovian dynamics, as it
cannot increase under completely positive maps \cite{fujiwara2001quantum,suzuki2016entanglement,laurenza2018channel}, and hence it can be used as a witness of non-Markovianity. Originally, introducing  a flow of QFI as $\mathcal{I}_{\varphi}(t)=\text{d}F_{\varphi}(t)/{\text{d}t}  $ \cite{lu2010quantum},  
it has been proposed that if $ \mathcal{I}_{\varphi}(t)>0 $ for some $ t $, then the time evolution is non-Markovian. Nevertheless, the  efficiency of the QFI flow to detect the non-Markovianity in various  scenarios has not been yet compared with the other faithful witnesses of the non-Markovianity. On the other hand, recently, the HSS flow 
$ {\text{d}HSS_{\varphi}(t)}/{\text{d}t}  $ has been proposed as
a faithful witness of non-Markovianity \cite{jahromi2020witnessing} in low dimensional systems. Therefore, our results also provide a sanity check of the QFI flow as a witness of the non-Markovianity. Moreover, because the QFI is  always contractive under Markovian dynamics,  our results provide a strong evidence for contractivity of the HSS under memoryless evolution of  high-dimensional systems  
  and pave the way to further studies on its applications in measuring the non-Markovianity in open quantum systems made of qudits.

\section*{Acknowledgements}

H.R.J. thanks Manuel Gessner for invaluable
comments as well as constructive remarks. H.R.J. also  wishes to acknowledge the financial support of the MSRT of Iran and Jahrom University.

\nocite{*}
\bibliography{Ref}% Produces the bibliography via BibTeX.

\end{document}